\documentclass[english,aps]{revtex4}
\usepackage{ae}
\usepackage{aecompl}
\usepackage[T1]{fontenc}
\usepackage[latin1]{inputenc}
\usepackage{graphicx}
\usepackage{amssymb}

\makeatletter

\providecommand{\tabularnewline}{\\}

\makeatletter

\makeatletter

\renewcommand{\[}{\begin{equation}}
\renewcommand{\]}{\end{equation}} 
\usepackage{hyperref}

\makeatother

\makeatother

\usepackage{babel}
\makeatother
\begin{document}

\preprint{This line only printed with preprint option}

\title{Lattice-switch Monte Carlo for binary hard-sphere crystals}

\author{A. N. Jackson}

\affiliation{SUPA, School of Physics, University of Edinburgh, Edinburgh EH9 3JZ,
Scotland, United Kingdom}

\author{G. J. Ackland}

\email{G.J.Ackland@ed.ac.uk}

\affiliation{SUPA, School of Physics, University of Edinburgh, Edinburgh EH9 3JZ,
Scotland, United Kingdom}

\begin{abstract}
We show how to generalize the Lattice Switch Monte Carlo method to
calculate the phase diagram of a binary system. A global coordinate
transformation is combined with a modification of particle diameters,
enabling the multi-component system in question to be explored and
directly compared to a suitable reference state in a single Monte
Carlo simulation. We use the method to evaluate the free energies
of binary hard sphere crystals. Calculations at moderate size ratios,
$\alpha=0.58$ and $\alpha=0.73$, are in agreement with previous
results, and confirm AB2 and AB13 as stable structures. We also find
that the AB(CsCl) structure is not entropically stable at the size
ratio and volume at which it has been reported experimentally, and
therefore that those observations cannot be explained by packing effects
alone.
\end{abstract}
\maketitle

\section{Introduction}

The most efficient packing of hard spheres is one of the most easily
posed questions in physics. For monodisperse spheres, it forms the
basis of Kepler's conjecture\cite{K1609:Keplar,H1998:KepSol}, and
the solution is any one of many degenerate close-packing arrangements.
In physical systems such as opals and colloidal suspensions, the problem
is generalised to calculating the highest entropy structure of a given
density. Away from the degenerate close-packing limit, face-centred
cubic turns out to be the most stable, and its exact entropy advantage
has been accurately calculated \cite{ADB1997:lattice-switch}. A further
generalization, which we consider here, is the packing of spheres
of different radii: the so-called binary hard sphere system.

In order to calculate stable packings for binary hard spheres, one
must evaluate the free energies of candidate phases as a function
of density, composition and size ratio, so that the entire phase diagram
can be constructed by minimization of the total free energy. 

A common approach to free energy calculation is use thermodynamic
integration. This involves the construction of a reversible path between
the structure of interest and some reference system. By performing
a simulation or set of simulations that traverse that path, one can
estimate the free energy difference between the reference and target
systems by integrating the derivative of the free energy along that
path. For example, the integration technique of Frenkel and Ladd \cite{FRENKEL1984}
uses an Einstein crystal as a reference system, and has been successfully
applied to a wide range of target systems, including binary hard sphere
solids \cite{ELDRIDGE1993,ELDRIDGE1993b,Eldridge:1993aa,FRENKEL1984}.
However, the integration process introduces an intrinsic source of
systematic error. Small systematic errors can accumulate at each step
in the integration, and integrating across the singularity at a phase
transition causes problems. Of course, careful application can minimise
these effects, but it remains preferable to have a technique where
systematic errors are more easily quantifiable. The most accurate
free energy differences in hard sphere systems are obtained from lattice-switch
Monte Carlo, \cite{ADB1997:lattice-switch} which allows two phases
to be sampled directly during a single Monte Carlo simulation, mapping
between them using a global coordinate transformation. Provided some
biased sampling method ensures that the phase transformation move
is accepted, the free-energy difference between phases can be determined
directly from the equilibrium probability of finding the simulation
in each phase. Finite size effects are the only source of systematic
errors. This approach was first used to study the solid state phase
behavior of monodisperse hard spheres \cite{ADB1997:lattice-switch,ABD2000:LSMCmethod}.
The method was subsequently developed for different energy models:
{}``Hamiltonian Switch'' and off-lattice: {}``phase-switch''.
It has been applied to the freezing transition for hard spheres in
\cite{NBW2000:phaseswitch,Wilding2002} and also to soft potentials
in \cite{Jackson:2002aa,errington:3130,McNeil-Watson2006}.

Here we investigate whether this methodology can be applied successfully
to multi-component systems by generalising the lattice-switch approach
to examine the structural phase behaviour of binary hard spheres.
The binary hard sphere system is well studied, and therefore the validity
of the generalised lattice switch method can be verified by direct
comparison with the results from previous experimental and theoretical
work.

Stable phases of the binary-hard sphere system are characterised primarily
by the diameter ratio, $\alpha=\sigma_{B}/\sigma_{A}$, where the
large and small particles are designated as A and B respectively.
At high values of $\alpha$, from $1.0$ down to $\approx0.92$, the
system is expected to form a substitutionally disordered crystal,
where A and B particles are randomly distributed on an FCC lattice
\cite{Barrat:1986aa,0022-3719-20-10-011}. At slightly lower values,
in the range $0.92\gtrsim\alpha\gtrsim0.85$, this mixed crystal becomes
metastable, and the solid phase phase separates into FCC crystals
of A and B particles \cite{Barrat:1986aa,0022-3719-20-10-011}.

\begin{figure}
\includegraphics[width=16cm]{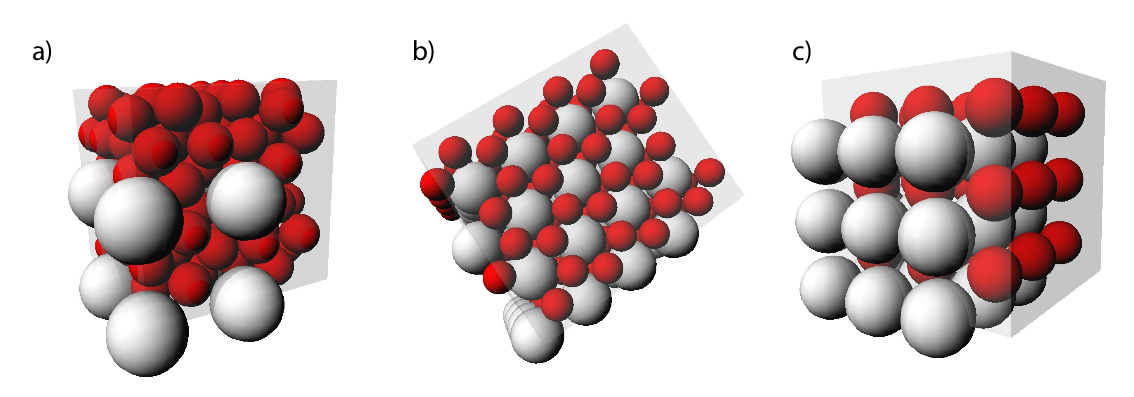}

\caption{\label{fig:structures}The three binary structures considered in
this work, with the A particles shown as pale spheres, B particles
shown as darker spheres, and the simulation cell shown as a translucent
box. a) The AB13 structure consists of a simple cubic lattice of large
spheres with icosahedral clusters of 13 B particles in the body-center
of each cell. The 112 particle unit cell contains 8 B-clusters arranged
in alternating orientations, where the clusters are rotate by $90^{\circ}$
between adjacent sub-cells. b) In AB2, hexagonal close-packed layers
of A particles are stacked directly on top of each other, with a honeycomb
pattern of B particles places in the interstitial sites between the
A layers. c) The AB(CsCl) structure, consisting of two interlaced
simple cubic lattices, with the B particles at the body-centre of
the A cube.}
\end{figure}

Below this, for $0.85\gtrsim\alpha\gtrsim0.5$, three different lattices
(shown in Figure \ref{fig:structures}) have been observed experimentally:
AB2, AB13 and AB(CsCl). AB13 and AB2 have been observed in Brazilian
gem opals (colloidal silica crystals) \cite{Sanders1980,Murray1980},
and later in poly-methylmethacylate (PMMA) colloidal dispersions,
both of which closely resemble hard spheres. In PMMA, AB13 was first
seen in \cite{BARTLETT1990} for $\alpha\approx0.61$, followed by
AB2 and AB13 at $\alpha\approx0.58$ in \cite{PUSEY1994}. These two
structures were determined to be unstable above $\alpha\approx0.62$
in \cite{ELDRIDGE1995} and below $\alpha\approx0.52$ in \cite{Hunt2000}.
More recent experiments have found AB2 to be stable for $0.60\gtrsim\alpha\gtrsim0.425$
and AB13 stable over $0.62\gtrsim\alpha\gtrsim0.485$ \cite{Schofield2005}.

These complex lattices have also seen in systems that bear little
resemblance to hard-spheres, such as rare gas solids \cite{Layer2006},
nanocrystals \cite{Redl:2003aa}, charge-stabilised colloids \cite{Royall2006},
and block copolymer micelles \cite{Abbas:2006aa}. At first, it is
surprising that these structures, particularly the highly complex
AB13 superlattice, should form spontaneously in such a wide range
of systems. However, it appears that entropic effects of hard-sphere
packing are sufficient to stabilise these structures.

Previously, the lattice switch approach has been used for the free
energy difference between two candidate structure for a single species.
Two-species phase behaviour is more complicated, as the stable state
of a binary system may consist of co-existing phases of different
compositions, at equal pressures and chemical potentials. Therefore,
instead of just comparing free energy of pairs of candidate phases,
we must to evaluate the absolute free energies of all competing phases
so that the overall phase diagram can be constructed. The basic approach
has been presented in \cite{BARTLETT1990a} by Bartlett, who considered
only the fluid and pure (face-centered cubic A or B) crystal phases.
We shall follow the work of Eldridge \textit{et al} \cite{ELDRIDGE1993,ELDRIDGE1993b,ELDRIDGE1995,Eldridge:1993aa}
and Cottin and Monson \cite{COTTIN1995,COTTIN1995a} and also include
AB2, AB13 and AB(CsCl) as candidate two-component crystals. In the
Eldridge \cite{ELDRIDGE1993} work, the stability of AB2 and AB13
for binary hard spheres was determined using the thermodynamic integration
method of Frenkel and Ladd \cite{FRENKEL1984}, with further details
of the AB2 calculations in \cite{ELDRIDGE1993b} and AB13 in \cite{Eldridge:1993aa}.
These results show general agreement with the experimental systems,
but suggest that AB2 is kinetically repressed at the AB2 1:2 stoichiometry,
as it only appears at higher concentrations of B particles \cite{Schofield2005}.

Theoretical studies of AB2, AB13 \& AB(CsCl) have been also carried
out via cell theory \cite{COTTIN1995,COTTIN1995a}, and density functional
theory (DFT) \cite{XU1992,DENTON1990,SMITHLINE1987}. For AB2 and
AB13, the results are generally consistent with both the integration
method and experimental results, although some minor differences remain.
In particular, the range of stability of AB13 is computed via cell
theory to be $0.61\ge\alpha\ge0.54$ \cite{COTTIN1995a}, whereas
the integration method finds AB13 to be stable over the range $0.626\ge\alpha\ge0.474$
\cite{Eldridge:1993aa}. Although the latter work does not consider
the possibility that AB2 is more stable than AB13 over this range
of $\alpha,$ it is notable that the range of stability determined
by integration methods is consistent with the range of observation
of AB13 found in the experimental work presented in \cite{Schofield2005}.

The case of the AB(CsCl) structure is somewhat less clear. The close-packed
density of this structure peaks at $\alpha=0.732$, and so may compete
with the slightly denser FCC structure, and indeed metastable CsCl
has been observed experimentally of at $\alpha=0.736$ in \cite{Schofield:2001bv}.
The stability of this structure has not been ascertained by integration
methods: it was not considered to be a valid candidate structure in
\cite{ELDRIDGE1995}. The cell theory work in \cite{COTTIN1995a}
did not find CsCl to be stable for any $\alpha$, but DFT results
have shown metastablity \cite{DENTON1990}, and even stablity \cite{SMITHLINE1987}.
However, the work in \cite{SMITHLINE1987} did not consider full phase
separation into coexisting A and B FCC crystals; they only consider
a fluid co-existing with a crystal phase made up of randomly distributed
A and B particles, but rich in A, and the fluid composition appears
to be fixed at 1:1. Interestingly, the AB(CsCl) structure has been
stabilised experimentally by using oppositely charged colloids \cite{Leunissen2005,Bartlett2005,Hynninen2006a,Royall2006}.
Moreover, BCC is the lowest energy arrangement for charged monodisperse
particles, so even if the AB(CsCl) structure is only metastable in
general, it is possible that a small amount of charge transfer between
particles or between particles and fluid is enough to stabilise the
structure.

For even smaller sphere size ratios ($\alpha\lesssim0.44$), the situation
becomes more complicated as additional structures start to appear
(e.g. AB(rocksalt) \cite{Trizac1997}) and because, at lower volume
fractions (or sufficiently low $\alpha$), the small spheres are free
to move from site to site within the crystal of larger spheres. Such
systems are not well suited to examination via the lattice-switch
approach, as the algorithm would have to be modified heavily in order
to cope in the free movement of small spheres. At smaller size ratios
still, the success of the perturbative approach \cite{Velasco1999}
suggests that it is best to model the presence of very small spheres
via a depletion potential.

Given the range of results that exist for binary hard spheres, it
provides an excellent testing ground for the extension of the lattice-switch
approach to multi-component systems. As our primary concern is simply
to extend the lattice-switch technique to the binary system, it makes
sense to choose a particular value of $\alpha$ to examine, instead
of trying to explore a whole range of diameter ratios before the correctness
of the approach has been determined. We have chosen to examine the
stability of AB2 and AB13 at $\alpha=0.58$, as there has been a large
amount of theoretical and experimental work published for that particular
diameter ratio. Furthermore, due to the ambiguity concerning the stability
of the AB(CsCl) structure, we shall also use the lattice-switch approach
to examine the stability of AB(CsCl) relative to the fluid phase and
phase-seperated solid state at $\alpha=0.73$.

\section{Formulation\label{sec:formulation}}

We consider a system of N particles, of spatial coordinates $\{\vec{r}\}$
and diameters $\{\sigma\},$ confined within a volume V, and subject
to periodic boundary conditions. The configurational energy for this
system of hard spheres has the form

\begin{equation}
E(\{\vec{r}\})=\left\{ \begin{array}{l}
0\textrm{ if }r_{ij}\leq\sigma_{ij}\;\forall\; i,j,\\
\infty\textrm{ otherwise, }\end{array}\right.\label{eq:EofRs}\end{equation}

where $r_{ij}=\left|\vec{r_{i}}-\vec{r_{j}}\right|$ and $\sigma_{ij}=\frac{1}{2}\left(\sigma_{i}+\sigma_{j}\right)$.
Each microstate has a Boltzmann weight $\exp[-\beta E]$, and so the
total configurational weight associated with this system can be written
as

\begin{equation}
\Omega(N,V)=\prod_{i}\left[\int_{V}d\vec{r_{i}}\right]\prod_{\langle ij\rangle}\Theta(r_{ij}-\sigma_{ij}),\end{equation}

where $\Theta(x)\equiv1(0)$ for $x\ge0(<0)$, and the product on
$\langle ij\rangle$ extends over all particle pairs. The associated
entropy density is

\begin{equation}
s(N,V)\equiv\frac{1}{N}\ln\Theta(N,V).\end{equation}

This formulation describes the total entropy, integrating over all
possible positions and so over all possible phases. In order to compare
the statistical weights of the configurations associated with individual
(candidate) phases, we must formulate a constraint that identifies
a configuration as `belonging to' a given phase. To do this, we decompose
the particle position coordinates into a sum of `lattice' and `displacement'
vectors:

\begin{equation}
\vec{r_{i}}=\vec{R_{i}^{\mathcal{A}}}+\vec{u_{i}},\label{eq:decomp}\end{equation}

where $\{\vec{R}\}_{\mathcal{A}}\equiv\vec{R_{i}^{\mathcal{A}},i=1,...,N}$
forms a set of fixed vectors associated with the crystal structure
labeled $\mathcal{A}$. This set is defined as the orthodox crystallographic
lattice convolved with the orthodox basis, but will be referred to
here as simply the `lattice vectors'. In previous work the coordinates
$r_{i}$ have denoted a pure phase arrangement of the particles. However,
it is possible for $\mathcal{A}$ to represent a two-phase mixture
or for particle sizes to be changed in the switch. We are thus able
to initialise a simulation within a particular phase, using the appropriate
set of lattice vectors ($\{\vec{R}\}_{\mathcal{A}}$) and diameters
($\{\sigma\}_{\mathcal{A}}$) for that phase, just by setting $\vec{u_{i}}=\vec{0}\;\forall\; i$.
We can now use a simple spatial criterion to decide whether a particular
configuration belongs to the same phase as the simulation progresses.
Periodically, throughout the simulation, the particles are checked
to determine whether, after having taken any centre-of-mass drift
into account, $\vec{u_{i}}$ is comparable to the interparticle spacing
(for more details, see section \ref{sub:Monte-Carlo-procedure}).
This can only happen if the particles have been rearranged and the
structure has been modified. so provides a robust criterion for ensuring
that a set of microstates belong to the same macrostate. We note that
this restriction of the configurational integral to pure perfect phases
erases contributions from equilibrium concentrations of lattice defects,
for example lattice vacancies, however for the systems in question
such defects are extremely rare and make no significant contribution
to total free energy \cite{Bowles:1994aa}.

By performing this coordinate decomposition (eq. \ref{eq:decomp}),
we have effectively changed our stochastic variables from being $\{\vec{r}\}$
to $\left[\{\vec{u}\},\mathcal{A}\right]$, operating under a modified
configuration energy function (c.f. eq. \ref{eq:EofRs}) of the form

\begin{equation}
E(\{\vec{u}\},\mathcal{A})=\left\{ \begin{array}{l}
0\textrm{ if }r_{ij}\leq\sigma_{ij}\;\forall\; i,j,\\
\infty\textrm{ otherwise, }\end{array}\right.\end{equation}

where the lattice label $\mathcal{A}$ dictates the choice of lattice
vectors via $r_{ij}(\{\vec{u}\},\mathcal{A})=\left|\vec{(R_{i}^{\mathcal{A}}}+\vec{u_{i}})-\vec{(R_{j}^{\mathcal{A}}}+\vec{u_{j}})\right|$
and diameters via $\sigma_{ij}(\mathcal{A})=\frac{1}{2}\left(\sigma_{i}^{\mathcal{A}}+\sigma_{j}^{\mathcal{A}}\right)$.
The configuration weight associated with a particular structure can
now be written as

\begin{equation}
\Omega(N,V,\mathcal{A})=\prod_{i}\left[\int_{\mathcal{A}}d\vec{r_{i}}\right]\prod_{\langle ij\rangle}\Theta(r_{ij}-\sigma_{ij}),\end{equation}

where $\int_{\mathcal{A}}$ signifies integration subject to the single-phase
constraint. The associated entropy density becomes

\begin{equation}
s(N,V,\mathcal{A})\equiv\frac{1}{N}\ln\Omega(N,V,\mathcal{A})\end{equation}

Once we construct a simulation capable of visiting the two different
phases ($\mathcal{A},\mathcal{B}$), the entropy difference between
them can be determined as:

\begin{equation}
\Delta s_{\mathcal{\mathcal{A}B}}\equiv s(N,V,\mathcal{A})-s(N,V,\mathcal{B})=\frac{1}{N}\ln\mathcal{R}_{\mathcal{A}\mathcal{B}}(N,V),\end{equation}

where

\begin{equation}
\mathcal{R}_{\mathcal{A}\mathcal{B}}(N,V)=\frac{\Omega(N,V,\mathcal{A})}{\Omega(N,V,\mathcal{B})}=\frac{P(\mathcal{A}|N,V)}{P(\mathcal{B}|N,V)}.\end{equation}

Here $P(\mathcal{A}|N,V)$ is the joint probability that a system
free to visit both structures will be found in the $\mathcal{A}$
structure.

\section{Implementation and Methodology}

\subsection{Defining the structures and the switch}

To evaluate the free energy of a particular binary crystal, we need
to perform a lattice switch simulation capable of visiting the target
phase, $\mathcal{A}$, and some reference phase, $\mathcal{B}$. As
an accurate semi-empirical expression for the free energy of a monodisperse
hard sphere FCC crystal has been determined by Alder \emph{et al}
\cite{ALDER1964,BJA1968:MDfccvhcpHS}, we choose this structure to
be our usual reference state. A lattice switch which kept the diameters
of the particles fixed (such that $\{\sigma\}_{\mathcal{A}}\equiv\{\sigma\}_{\mathcal{B}}$),
would require either a single simulation box containing co-existing
crystals of A and B particles (and thus associated interfacial effects,
making for a poor reference system) or two separate boxes corresponding
to separate A and B crystals. This latter approach can be made to
work, but the complexity associated with adding a second simulation
cell is unnecessary when the two cells are being used to simulate
two \emph{identical} structures that differ \emph{only} in scale (by
$\alpha)$. A simpler option is to allow the diameters to change during
the switch, turning all $\sigma_{B}$ particles in the mixed phase
into particles of diameter $\sigma_{A}$, allocated different sites
on a single shared FCC lattice. This creates a switch between a binary
crystal and a single crystal, and ensures that the number of degrees
of freedom in each system is the same. Furthermore, as the free-energy
difference between FCC and HCP is accurately known \cite{ABD2000:LSMCmethod},
we are free to use either close-packed structure as a reference crystal,
which affords us a little more flexibility when it comes to choosing
system sizes. The system sizes used are shown in Table \ref{tab:systems}.

\begin{table}
\begin{tabular}{|c|c|c||c|c||c|c|c|}
\hline 
\multicolumn{1}{|p{1.2cm}|}{}&
\multicolumn{2}{c||}{\textbf{Target Crystal}}&
\multicolumn{2}{c||}{\textbf{Reference Crystal}}&
Runtime&
Runtime&
Runtime\tabularnewline
\hline 
\textbf{N}&
Structure&
Dimensions&
Structure&
Dimensions&
MCS {[}EOS]&
MCS {[}WF]&
MCS {[}DF]\tabularnewline
\hline
\hline 
112&
AB13&
$1\times1\times1$&
HCP&
$4\times7\times4$&
$5\times10^{5}$&
$10^{6}\;[40]$&
$4\times10^{6}[8]$\tabularnewline
\hline 
896&
AB13&
$2\times2\times2$&
FCC&
$8\times8\times14$&
$5\times10^{5}$&
$6\times10^{6}\;[240]$&
$6\times10^{6}[10]$\tabularnewline
\hline 
192&
AB2&
$4\times4\times4$&
FCC&
$8\times4\times6$&
$10^{6}$&
$10^{6}\;[40]$&
$4\times10^{6}\;[80]$\tabularnewline
\hline 
648&
AB2&
$6\times6\times6$&
FCC&
$8\times9\times9$&
$10^{6}$&
$5\times10^{6}\;[200]$&
$2\times10^{6}\;[200]$\tabularnewline
\hline 
54&
CsCl&
$3\times3\times3$&
FCC&
$6\times3\times3$&
$5\times10^{5}$&
$10^{6}\;[40]$&
$4\times10^{6}\;[40]$\tabularnewline
\hline 
128&
CsCl&
$4\times4\times4$&
HCP&
$8\times4\times4$&
$5\times10^{5}$&
$2\times10^{6}\;[80]$&
$5\times10^{6}\;[20]$\tabularnewline
\hline 
432&
CsCl&
$6\times6\times6$&
FCC&
$8\times6\times9$&
$5\times10^{5}$&
$3\times10^{6}\;[120]$&
$8\times10^{6}\;[40]$\tabularnewline
\hline
\end{tabular}

\caption{\label{tab:systems}Structures and sizes used in this work. For FCC,
HCP and AB2, the unit-cells were oriented so that the close-packed
layers were lying in the x-y plane. The cubic cells of AB13 and CsCl
were made commensurate with the simulation cell. The run times are
shown for determining the equation of state (EOS), the weight function
(WF) and the free-energy difference (DF), in units of Monte Carlo
sweeps (MCS). Numbers in square brackets indicates the largest number
of blocks the run was split into for block analysis. For all structures
and sizes, 20,000 MCS runs were used to equilibrate the constant pressure
runs, while the constant volume simulations needed just 5,000 MCS
for equilibration.}
\end{table}

Once the crystal lattices have been chosen, one must choose a site
mapping between crystals $\mathcal{A}$ and $\mathcal{B}$ . In the
previous lattice-switch work for FCC and HCP structures more efficient
switches which preserved close-packed planes were used. However, for
binary-to-single-lattice switches, there are no such obvious similarities,
therefore, we mapped sites between crystals at random.

To predict the equilibrium phase behaviour, we must first determine
the Gibbs free energy of all the candidate phases, as a function of
the pressure, $P^{\star}=P\sigma_{A}^{3}/kT$, and the composition,
$X=N_{B}/(N_{A}+N_{B})$, where $N_{A}$ and $N_{B}$ are the numbers
of A and B particles respectively. Naturally, we would seek to measure
the Gibbs free energy directly via a lattice-switch simulation carried
out in the constant pressure ensemble. However, when attempting to
switch between the pure and the binary crystals using a single simulation
cell, the fact that the equilibrium cell size is very different for
these two structures means that the switch becomes ineffective. The
lattice-switch only updates the lattice and the diameters, not the
cell parameters, and therefore performing the switch from the reference
lattice involves attempting to transform an equilibrium density pure
crystal into a loosely packed binary crystal (as reducing the diameter
of so many particles vastly increases the available volume). The conjugate
move requires a transformation from an equilibrium density binary
crystal to a pure crystal at such a high density that it is likely
that all particles overlap. To facilitate either switching move, the
volume of the simulation cell would have to be dilated so far that
the system would rapidly melt, and indeed such a switch cannot be
made to work in practice. It is possible, at least in principle, to
overcome this limitation by adding an explicit volume dilation to
the switch, so that the cell dimensions, the lattice and the diameters
are all changed simultaneously. However, adding a volume transformation
factor affects the probability of sampling the two phases in a non-trivial
way. For example, a dilation to a large cell will almost always be
accepted, whereas the compression to a small cell will be difficult
to achieve. This will bias the measured free energy difference by
a factor which could be determined by careful application of the mapping
transformation matrix approach outlined in \cite{ABD2000:LSMCmethod},
but this introduces a number of computational overheads, and the slow
switching rate mitigates against gathering good statistics.

All these issues can be avoided in the constant volume ensemble by
keeping the density fixed when switching between the reference and
target structures, thus ensuring that the displacements are \emph{reasonable}
in both phases because the distances between particles are \emph{comparable}
in both phases under these conditions.

\subsection{Simulation Methodology\label{sub:Simulation-Methodology}}

To estimate the Gibbs free energy, three simulations were carried
out for each state point. The first was a single phase constant pressure
(NPT) simulation, used to estimate the equilibrium density and cell
parameters (c/a ratio) for each binary crystal, at each pressure.
Then, two lattice-switch NVT simulations were performed at the volume
fractions and c/a ratios determined from the NPT simulations. The
lattice switch attempts to transform between the target crystal (at
the target density and c/a ratio) and a pure FCC crystal (at the same
density, but with c/a = 1.0). The first run is used to evaluate the
multicanonical weight function and the second uses this function to
determine the free energy difference between the crystals (see section
\ref{sub:Monte-Carlo-procedure} for more details).

Having determined the cell proportions, the density, and the Helmholtz
free energy difference between the two structures ($\Delta F$) for
a range of pressures, we can then build the absolute free energy of
the binary crystal by adding in the free energy of the pure phase.
To determine this, we integrate the Alder equation of state \cite{ALDER1964,BJA1968:MDfccvhcpHS,YOUNG1979},

\begin{equation}
\frac{PV}{NkT}=\frac{3}{V^{\star}-1}+2.566+0.55(V^{\star}-1)-1.19(V^{\star}-1)^{2}+5.95(V^{\star}-1)^{3},\label{eq:Zalder}\end{equation}

with respect to the reduced volume $V^{\star}=V/V_{0}$ (where $V_{0}$
is the close packed volume, $N\sigma^{3}/\sqrt{2}$). This yields
the Helmholtz free-energy:

\begin{equation}
\frac{F_{ex}^{Alder}}{NkT}=-3\ln\left(\frac{V^{\star}-1}{V^{\star}}\right)+5.124\ln V^{\star}-20.78V^{\star}+\frac{19.04}{2}V^{\star2}-\frac{5.95}{3}V^{\star3}+C_{4}+1,\label{eq:Falder}\end{equation}

where, following \cite{YOUNG1979}, the constant of integration $C_{4}$
is set to $15.05$. The Gibbs free energy can be constructed from
the Helmholtz form via:

\begin{equation}
G_{ex}=F_{ex}^{Alder}+\Delta F+PV-1-\ln\left[\frac{PV}{NkT}\right]\label{eq:FexToGex}\end{equation}

where the last two terms arise from the difference between the Gibbs
and Helmholtz free energies for an ideal gas. To compute the phase
diagrams based on these free energies, we follow the approach of Bartlett
\cite{BARTLETT1990} and assume that the particles are immiscible
in the solid phases, and that the binary fluid behaviour can be described
accurately by the Mansoori equation \cite{MANSOORI1971}. At each
pressure, the densities of the candidate phases are determined from
the equations of state. From these densities, the Gibbs free energies
and chemical potentials of the competing phases can be determined
for each constant pressure line. The condition of equal chemical potentials
is then applied along the constant pressure line, where the the coexistence
between the crystal phases and the Mansoori fluid is estimated from
the chemical potentials of each species in the fluid via (see \cite{COTTIN1995a}):

\begin{equation}
\mu_{A}^{f}+n\mu_{B}^{f}=(1+n)\times G(AB_{n}),\end{equation}

where $\mu_{A}^{f}$ and $\mu_{B}^{f}$are the chemical potentials
determined from the Mansoori fluid for the A and B particles respectively,
and $G(AB_{n})$ is the Gibbs free energy of the solid phase that
has $n$ B particles for each A particle.

\subsection{Multicanonical Biassed Monte Carlo procedure\label{sub:Monte-Carlo-procedure}}

In general, a lattice switch move from a zero-overlap microstate in
one phase is unlikely to map onto a zero-overlap microstate in the
other. Therefore, the simulation must monitor not only which phase
the is currently being explored, but also how close the simulation
is to being able to perform the lattice switch. To find a suitable
order parameter, we first let $M(\{\vec{u}\},\mathcal{A})$ denote
the number of overlapping pairs of particles associated with the displacements
$\{\vec{u}\}$, within the structure $\mathcal{A}$. We can then define
the overlap order parameter:

\begin{equation}
\mathcal{M}(\{\vec{u}\})\equiv M(\{\vec{u}\},\mathcal{B})-M(\{\vec{u}\},\mathcal{A}),\end{equation}

where $\mathcal{M}$ is necessarily $\ge0$ ($\le0$) for realizable
configurations of the $\mathcal{A}$ ($\mathcal{B}$) structure. To
ensure that the simulation can reach the so-called gateway states
($\mathcal{M}=0$), we apply the multi-canonical Monte Carlo method
\cite{BAB1999:mcmcrev} as a way of biasing the simulation in a controlled
fashion. To this end, we augment the system energy function such that

\begin{equation}
\mathcal{E}(\{\vec{u}\},\mathcal{A})=E(\{\vec{u}\},\mathcal{A})+\eta(\mathcal{M}(\{\vec{u}\})),\end{equation}

where $\eta(\mathcal{M}),\mathcal{M}=0,\pm1,\pm2,...$ constitute
a set of multicanonical weights \cite{BAB1999:mcmcrev}. Given that
a suitable weight function can be determined (see below), the canonical
distribution $P(\mathcal{M}|N,V)$ can be estimated from the measured
multicanonical distribution $P(\mathcal{M}|N,V,\{\eta\})$ with the
identification

\begin{equation}
\mathcal{R}_{\mathcal{AB}}(N,V)=\frac{\sum_{\mathcal{M}>0}P(\mathcal{M}|N,V)}{\sum_{\mathcal{M}<0}P(\mathcal{M}|N,V)}=\frac{\sum_{\mathcal{M}>0}P(\mathcal{M}|N,V,\{\eta\})e^{\eta(\mathcal{M})}}{\sum_{\mathcal{M}<0}P(\mathcal{M}|N,V,\{\eta\})e^{\eta(\mathcal{M})}}.\end{equation}

Here, the exponential reweighting maps the multicanonical distribution
onto the canonical, unfolding the bias associated with the weights.

The multicanonical weights are determined from non-equilibrium calculations,
which are fast and give a fair approximation to the equilibrium situation.
We conduct a series of separate Monte Carlo simulations, where the
particle displacements (and therefore $\mathcal{M}$) are reset to
zero at the start of each. As these simulations will each relax to
the equilibrium $\mathcal{M}$ values, this forces the overall run
to explore the full range of $\mathcal{M}$. Throughout this sequence
of simulation, the program monitors and accumulates the macrostate
transition probability matrix, as described in \cite{Smith:1996aa},
and at the end of the whole run, an estimate of the weight function
is determined from the transitions probabilities by a simple shooting
method \cite{Jackson2001}. This weight function is then used for
the second NVT run, during which the time spent in each phase is recorded
(with the biasing taken into account), and automatically block-analysed
to produce an estimate of the free-energy difference. The actual timings
varied depending on the target system and the number of particles,
and are shown in Table \ref{tab:systems}.

The simulations operated under the usual Metropolis algorithm \cite{1953NM:MC},
using random numbers generated by the Mersenne Twister algorithm \cite{272995},
as implemented in RngPack, version 1.1a \cite{Houle2003}. For each
move, a sphere is chosen at random, and a trial displacement is generated
by taking the displacement of that particle $\vec{u_{i}}$ and adding
a random three-dimensional vector $\Delta\vec{u_{i}}$ drawn uniformly
from a cube of fixed size. The update in the displacement is accepted
according to the usual Metropolis prescription:

\begin{equation}
p_{a}(\{\vec{u}\}\rightarrow\{\vec{u}\}+\Delta\vec{u_{i}})=\min\{1,\exp[-(\mathcal{E}(\{\vec{u}+\Delta\vec{u_{i}}\},\mathcal{A})-\mathcal{E}(\{\vec{u}\},\mathcal{A}))]\}.\end{equation}

Therefore, moves that would lead to overlaps in the current phase
are always rejected, and moves that would alter the number of overlaps
in the conjugate phase are accepted or rejected based on the multicanonical
weights. The magnitude of $\Delta\vec{u_{i}}$ was automatically adjusted
to yield an acceptance rate of about one third, as this has been shown
previously to produce an acceptable rate of decorrelation in $\mathcal{M}$
\cite{Jackson2001}. When $\mathcal{M}=0$, the lattice-switch move
is attempted once per Monte Carlo Sweep (MCS) (i.e. once every N particle
moves) on average, and is always accepted as there is no energy cost
or multicanonical weight difference associated with this move.

Due to the short-range nature of the hard sphere interaction, the
calculation of $E(\{\vec{u}\},\mathcal{A})$ does not need to extend
over all $N^{2}$ particle pairs. To make the simulation more efficient,
each particle only checks for interactions with its neighbours, up
to a maximum interaction distance of 1.5 unit cells (i.e. 1.5 times
the A-A separation). There are two lists of neighbours for each particle:
one for each phase. As mentioned in section \ref{sec:formulation},
we require a criterion to determine whether structure is stable, and
the simulation achieves this by periodically scanning the system for
particles that have moved further than the near neighbour distance
away from their site (after having taken any centre of mass diffusion
into account). If this is observed to happen at any point during any
simulation at a given pressure, then the structure is considered unstable,
and no longer considered as a candidate structure at that pressure.

To establish the equations of state for these systems, a number of
constant pressure simulations were also performed. In this case, the
simulation cell parameters (the lengths of the sides of the cell,
and therefore the volume) can fluctuate during the simulation. Such
moves are accepted with probability:

\begin{equation}
p_{a}(V\rightarrow V')=\min\{1,\exp[-(\mathcal{E}(\{\vec{u}\}',\mathcal{A})-\mathcal{E}(\{\vec{u}\},\mathcal{A}))-P^{\star}\Delta V+N\ln(V'/V)]\},\end{equation}

where $V'$ is the volume associated with the trial move. Of course,
this cell-size dilation can change $\mathcal{E}$ globally, and so
an expensive energy calculation must be carried out for each trial
volume move. Volume moves were attempted at random, once per MCS on
average. The maximum size of the random volume-dilation is automatically
adjusted so that the acceptance rate for volume moves is about one
half, which produces an acceptable rate of exploration of the volume
parameter space \cite{Jackson2001}.

\section{Results}

\subsection{Pressure curves and stability}

\begin{figure}
\includegraphics[width=10cm]{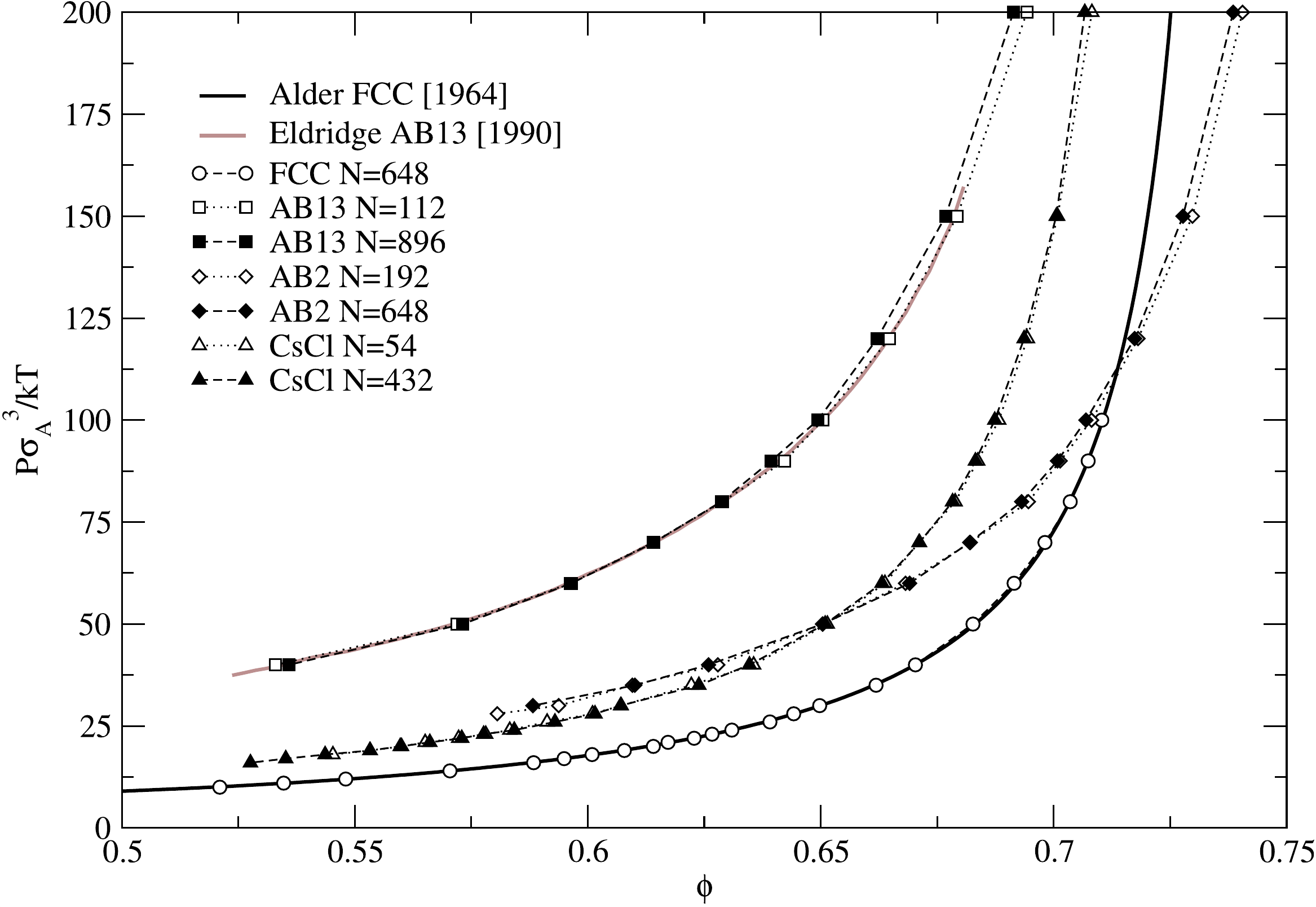}

\caption{\label{fig:eos}Equations of state determined in the constant pressure
ensemble, measuring the density as a function of the pressure. Errors
in the density estimates are smaller than the symbol size. The Alder
equation of state for the pure crystal \cite{ALDER1964} is shown
as a solid line, and the equation of state for AB13 as observed in
\cite{Eldridge:1993aa} is shown as a solid gray line.}
\end{figure}

The equations of state, as measured by Monte Carlo simulation in the
constant pressure ensemble, are shown in Figure \ref{fig:eos}. The
agreement with the data presented in \cite{Eldridge:1993aa} indicates
that the Molecular Dynamics simulation presented there used the smaller
($N=112$) system size when computing the equation of state. Also,
the close agreement between the Alder equation of state \cite{ALDER1964}
and the simulated FCC crystal is clear. However, we note that while
the Alder form (eq. \ref{eq:Zalder}) provides a reliable estimate
of the equation of state, the individual terms are difficult to reproduce.
Despite having determined the density of the FCC structure at each
pressure to within at least $0.1\%$, attempting to fit an equation
of the Alder form to this data led to a poorly-determined fit that
only reproduced the original terms to within one significant figure.
It is therefore clear that a much larger range of pressures must be
used to determine Alder-style equations of state for these structures,
and given that quite a modest range of pressures was explored in the
original work \cite{BJA1968:MDfccvhcpHS}, it is not clear that the
parameterization supplied in eq. \ref{eq:Zalder} is a particularly
accurate one.

\begin{figure}
\includegraphics[width=10cm]{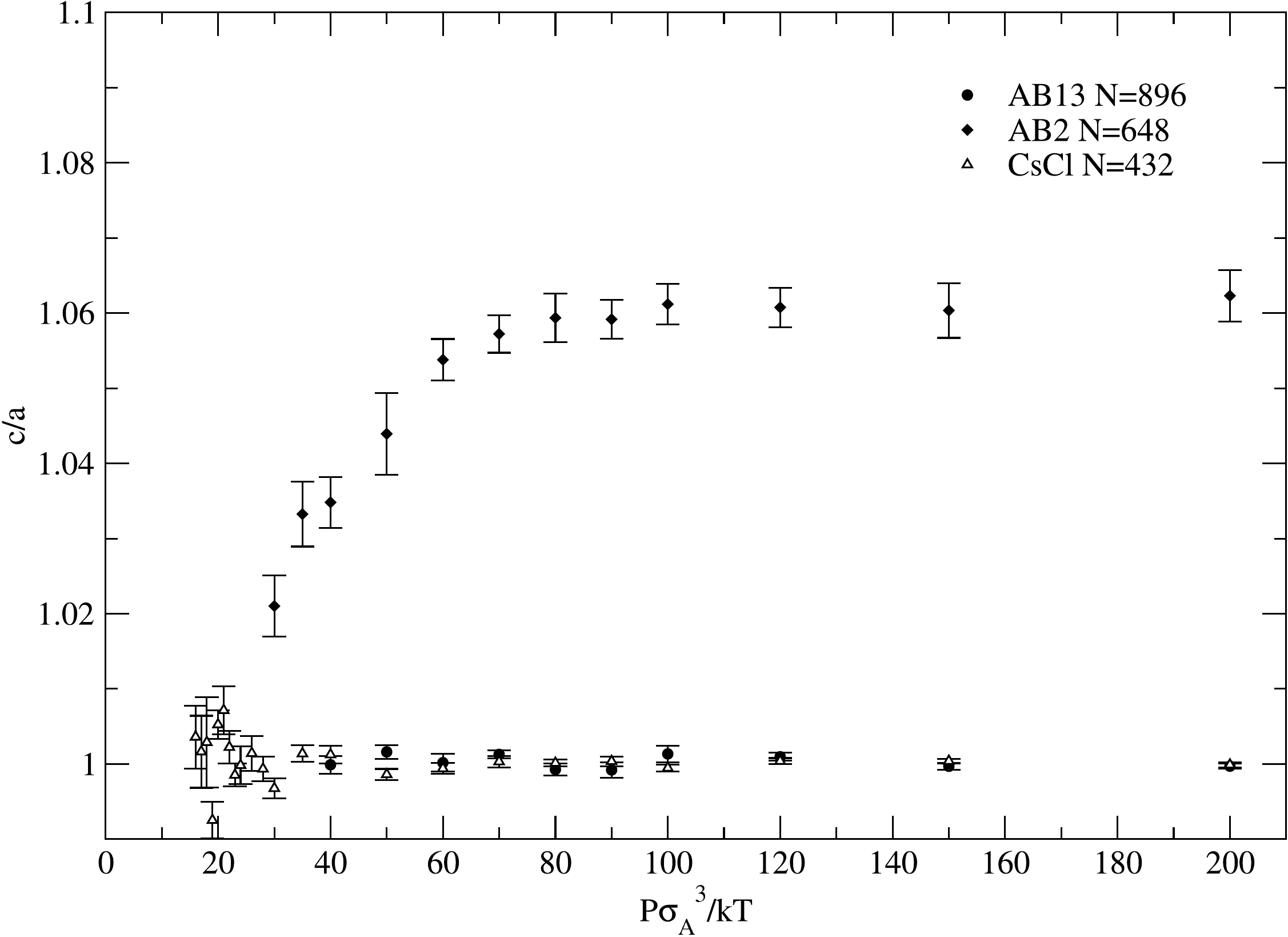}

\caption{\label{fig:cOa}Observed $c/a$ ratios for the target phases.}
\end{figure}

During the equation of state simulations, the cell proportions were
also monitored and Figure \ref{fig:cOa} shows the results in terms
of the c/a ratio. As expected on symmetry grounds, the AB13 and CsCl
crystals are stable at $c/a=1$. However, the AB2 crystal tends to
expand in the direction perpendicular to the alternating stacking
planes of A and B particles, and the degree of expansion found here
is consistent with the results presented in \cite{ELDRIDGE1993b}.
The c/a ratio decreases as the pressure is lowered towards the fluid
region, and as the simulations approach the melting point, the fluctuations
in the volume and aspect ratio of all the crystals become large as
the structures start to break down. As noted earlier, this break-down
condition discounts extended metastability and the possibility that
the structure may be stabilised by some equilbrium concentration of
defects, and so provides an upper bound on the actual limit of stability
of the structure.

\subsection{Free energies}

\begin{figure}
\includegraphics[width=10cm]{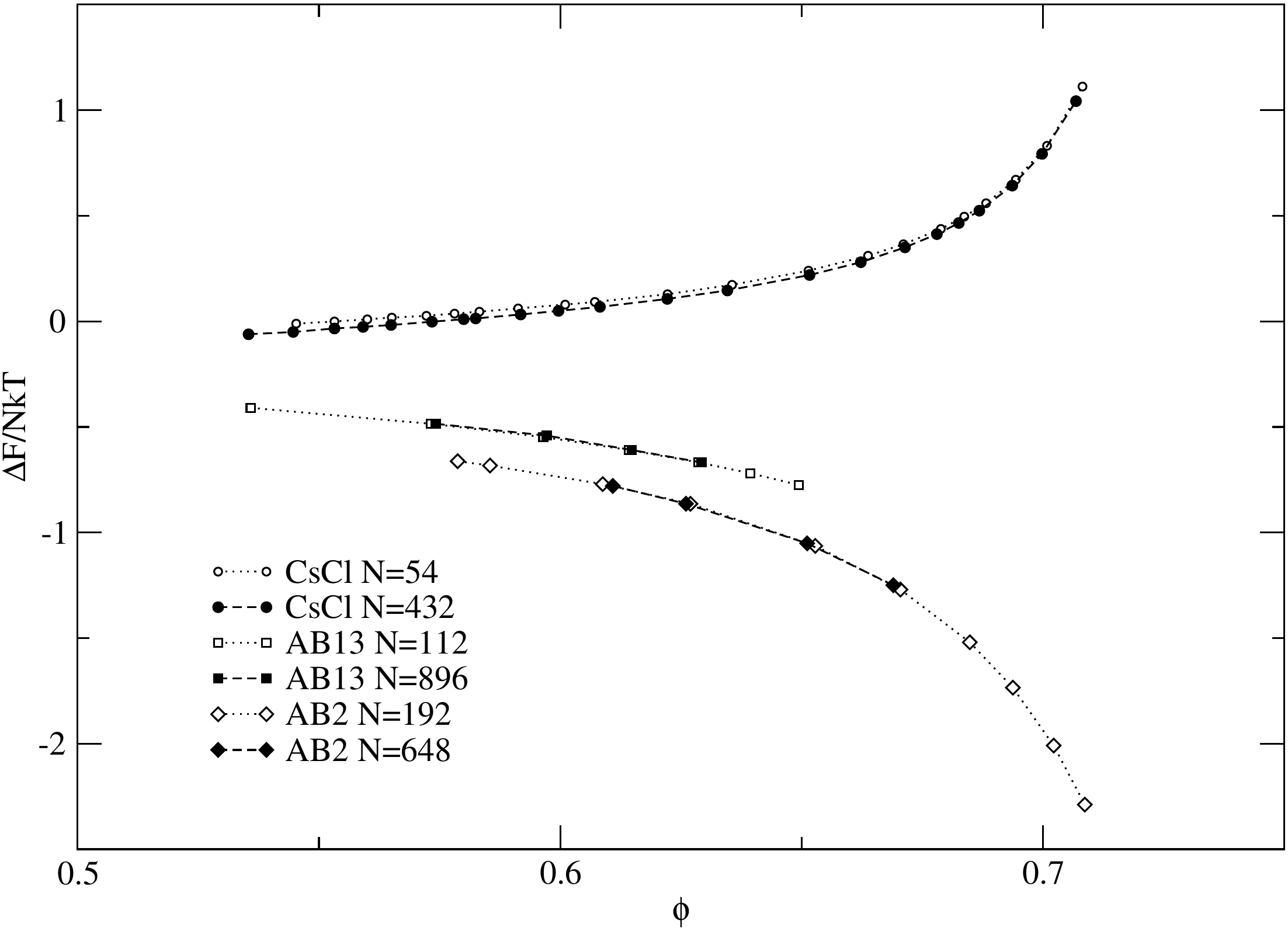}

\caption{\label{fig:AB13_AB2_Df}The raw Helmholtz free energy difference
per particle between the reference FCC and target AB2 and AB13 crystals
at $\alpha=0.58$, and AB(CsCl) at $\alpha=0.73$, as measured by
the lattice switch simulations. The final data points at low $\phi$
indicate the lowest density at which the structure was determined
to be stable. The points at the high density end indicate the upper
limit of the chosen pressure range explored by the simulations. Some
simulations were performed at larger system sizes, as shown here,
and indicated that the overall form of the free energy difference
does not depend strongly on the system size.}
\end{figure}

The free energy difference between the reference (FCC or HCP) and
target crystals AB2 and AB13 at $\alpha=0.58$, and for AB(CsCl) at
$\alpha=0.73$ are shown in Figure \ref{fig:AB13_AB2_Df}. In all
cases, the free energy difference between FCC and HCP $(\approx1\times10^{-3}kT$
per particle) is so small that it cannot be resolved on the scale
of these plots. The AB2 and AB13 structures are lower in free energy
than the FCC crystals for all densities at these size ratios.

\begin{figure}
\includegraphics[width=10cm]{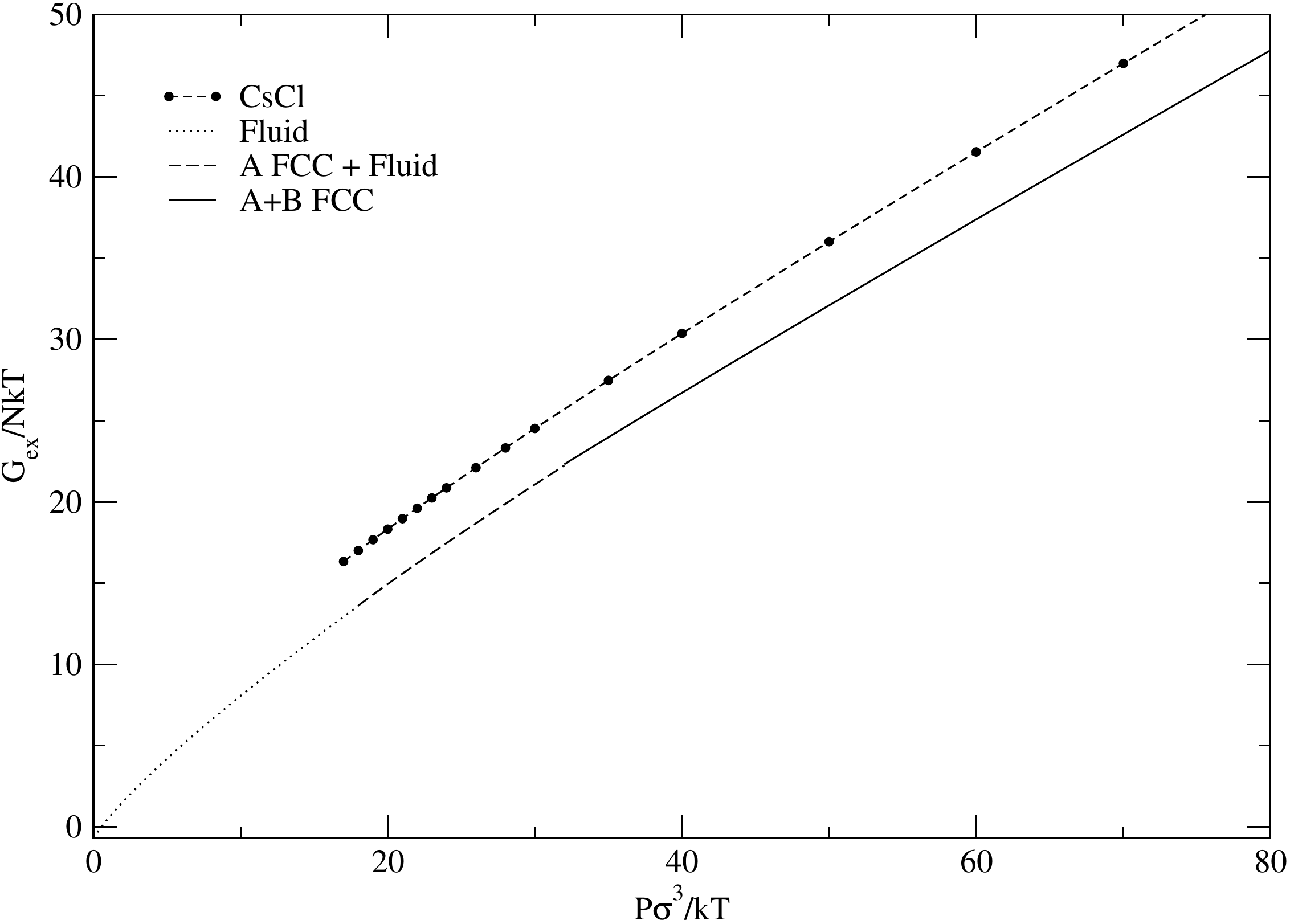}

\caption{\label{fig:Gibbs-CsCl}Gibbs free energies as a function of pressure
for AB(CsCl) $\alpha=0.72$ at a fixed composition of $X=0.5$. The
Gibbs free energies of the Mansoori fluid, the pure A FCC crystal
coexisting with a Mansoori fluid, and the dual FCC A and B crystal
phases are also shown. Finite size effects cannot be resolved on this
plot.}
\end{figure}

Figure \ref{fig:AB13_AB2_Df} also shows that the AB(CsCl) structure
has a significantly higher Helmholtz free energy than FCC at high
densities, the difference falls rapidly as the density is lowered
towards melting. However, it is important to remember that the AB(CsCl)
structure is not competing with a pure FCC crystal, but with one or
more coexisting phases at equal pressures and chemical potentials.
Figure \ref{fig:Gibbs-CsCl} illustrates this by comparing the Gibbs
free energy of AB(CsCl) with those of the competing phases - a Mansoori
fluid, a pure FCC crystal coexisting with a B-rich Mansoori fluid,
and a dual crystal state made up of separate A and B FCC crystals.
The AB(CsCl) structure is clearly not stable over this range, and
this result compares favourably with results from DFT calculations
in \cite{DENTON1990} and via cell theory in \cite{COTTIN1995a}.
However, we note that the free energy difference is quite small, and
so a small amount of charge asymmetry may be sufficient to stabilise
the structure in the low density solid phase.

\begin{figure}
\includegraphics[width=10cm]{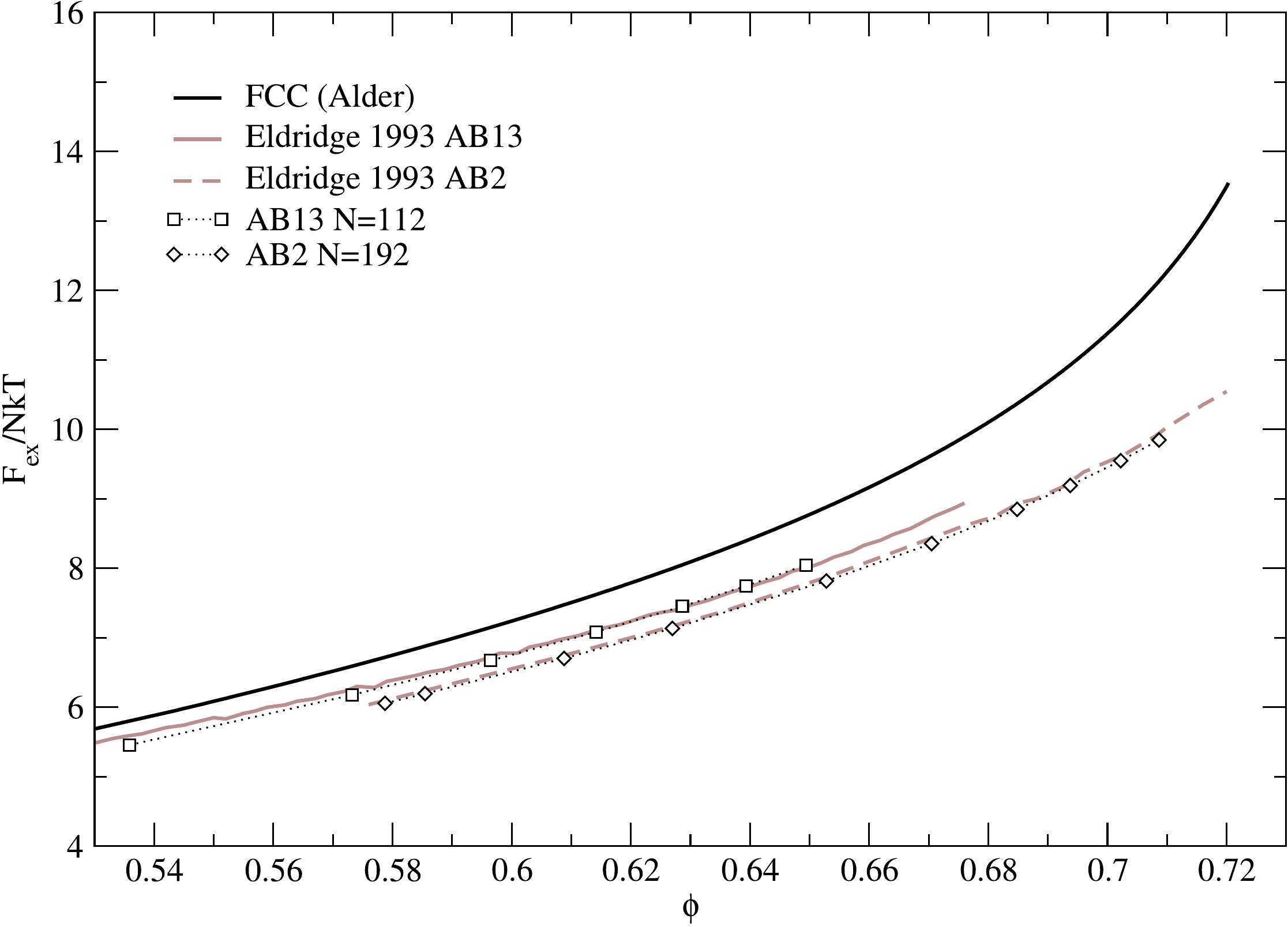}

\caption{\label{fig:FvVf}The excess (i.e. with respect to an ideal gas) Helmholtz
free energies of AB2, AB13 and FCC at $\alpha=0.58.$ Data from literature
for AB2 \cite{ELDRIDGE1993b} and AB13 \cite{Eldridge:1993aa} is
also shown for comparison.}
\end{figure}

\begin{figure}
\includegraphics[width=10cm]{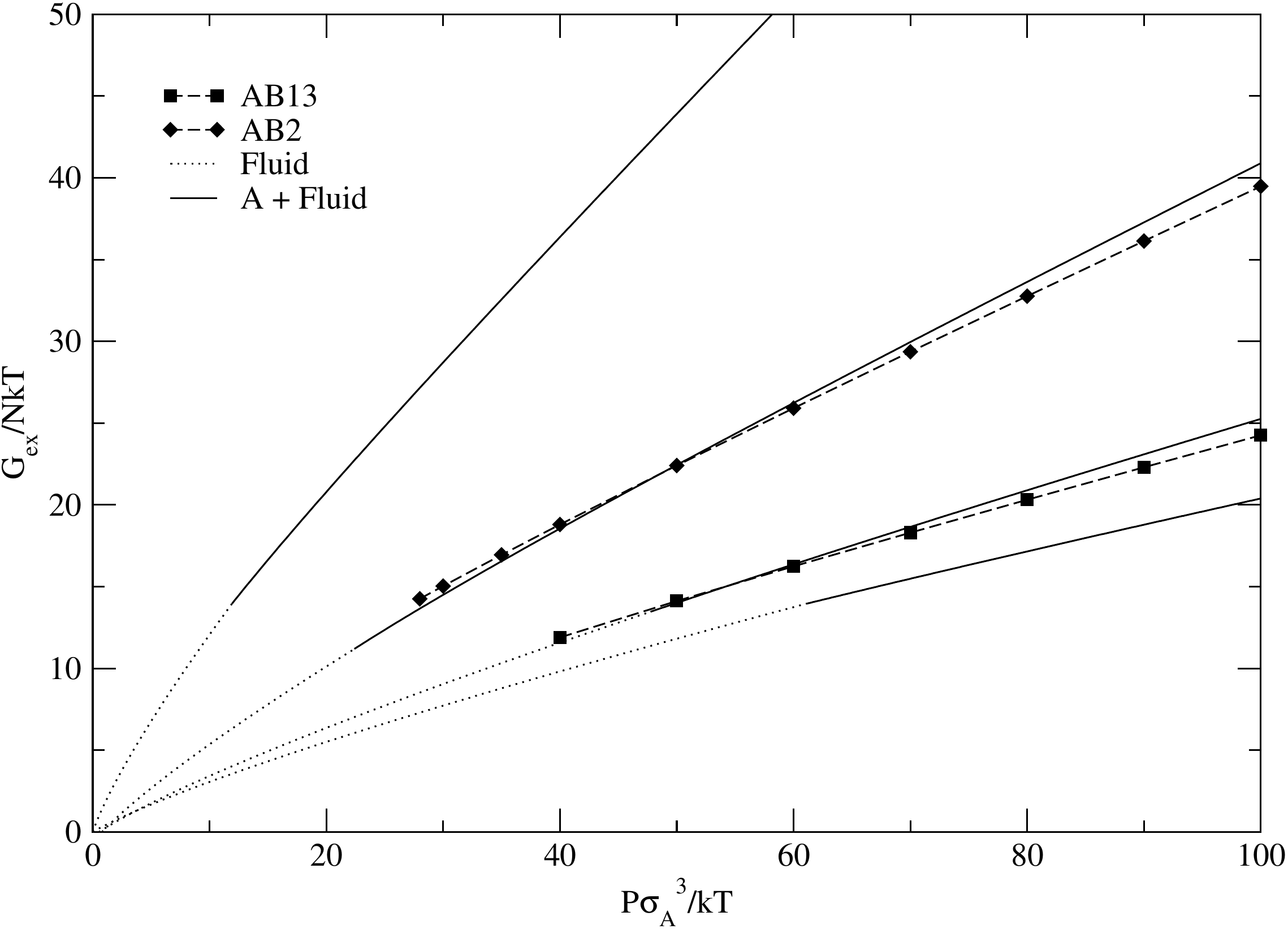}

\caption{\label{fig:GvPres}Gibbs free energies for AB2 and AB13 at $\alpha=0.58$.
The Gibbs free energies of Mansoori fluid and the A FCC crystal coexisting
with a B-rich Mansoori fluid at the appropriate compositions are also
shown. Finite size effects cannot be resolved on this plot.}
\end{figure}

For all structures, the free energy difference measured in the constant
density ensemble shows very little dependence on system size, and
even rather small systems provide reasonable estimates of the free
energy difference with respect to FCC. To allow comparison with the
literature results, the Helmholtz free energies of the structures
for $\alpha=0.58$ are shown in figure \ref{fig:FvVf}, and the Gibbs
free energies (computed via eq. \ref{eq:FexToGex}) are shown in Figure
\ref{fig:GvPres}. The lattice switch results are in clear agreement
with the results determined via integration methods \cite{Eldridge:1993aa,ELDRIDGE1993b},
and with the DFT results presented in \cite{XU1992} and the cell
theory results in \cite{COTTIN1995a}. For clarity, Figure \ref{fig:FvVf}
only includes the results presented in \cite{Eldridge:1993aa,ELDRIDGE1993b}
and omits the almost identical results of \cite{COTTIN1995,COTTIN1995a}.

\subsection{Phase diagrams}

\begin{figure}
\includegraphics[width=10cm]{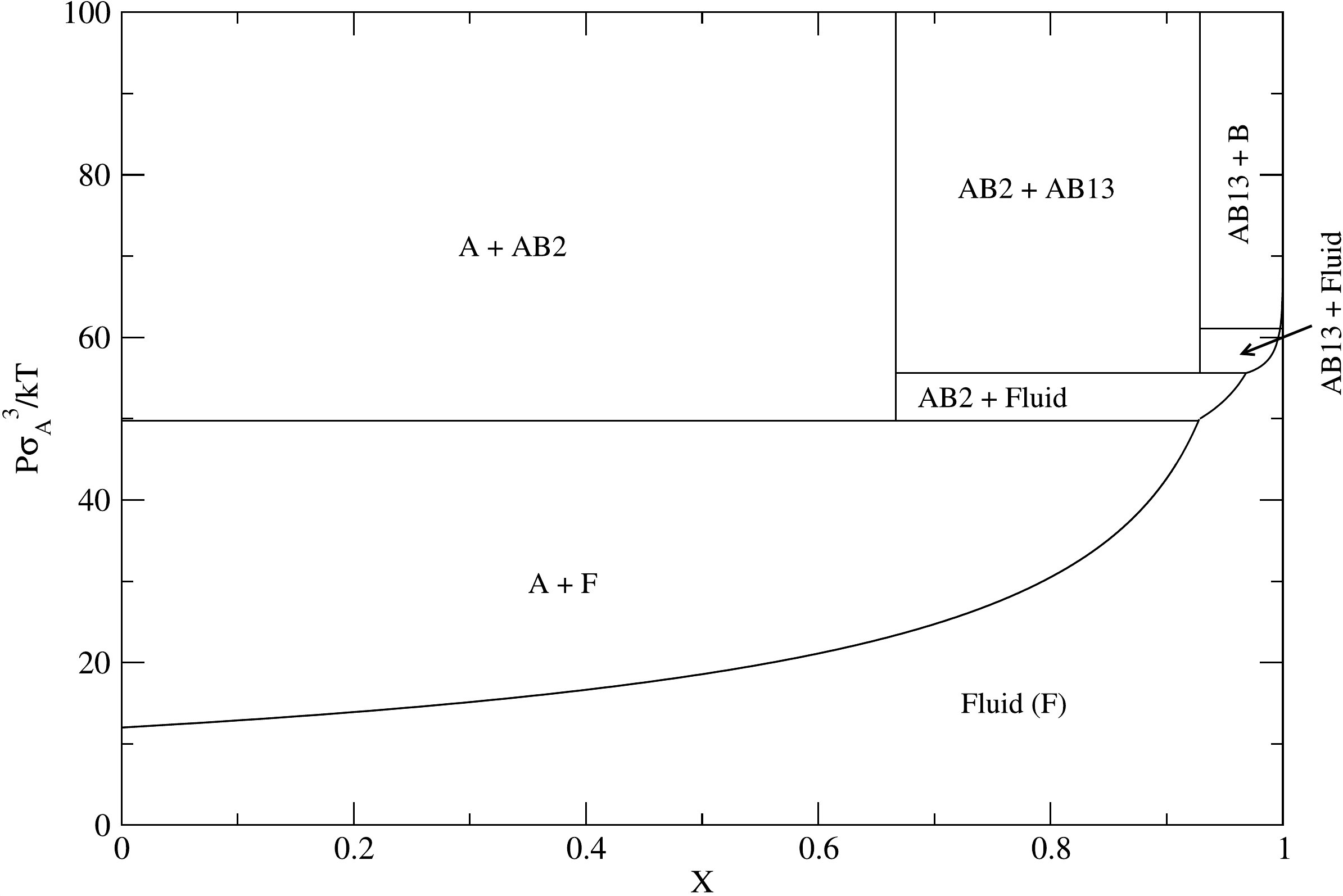}

\caption{\label{fig:PPhases}P-X Phase Diagram at $\alpha=0.58$, determined
from the lattice switch results using the methodology described in
section \ref{sub:Simulation-Methodology}.}
\end{figure}

\begin{figure}
\includegraphics[width=8cm]{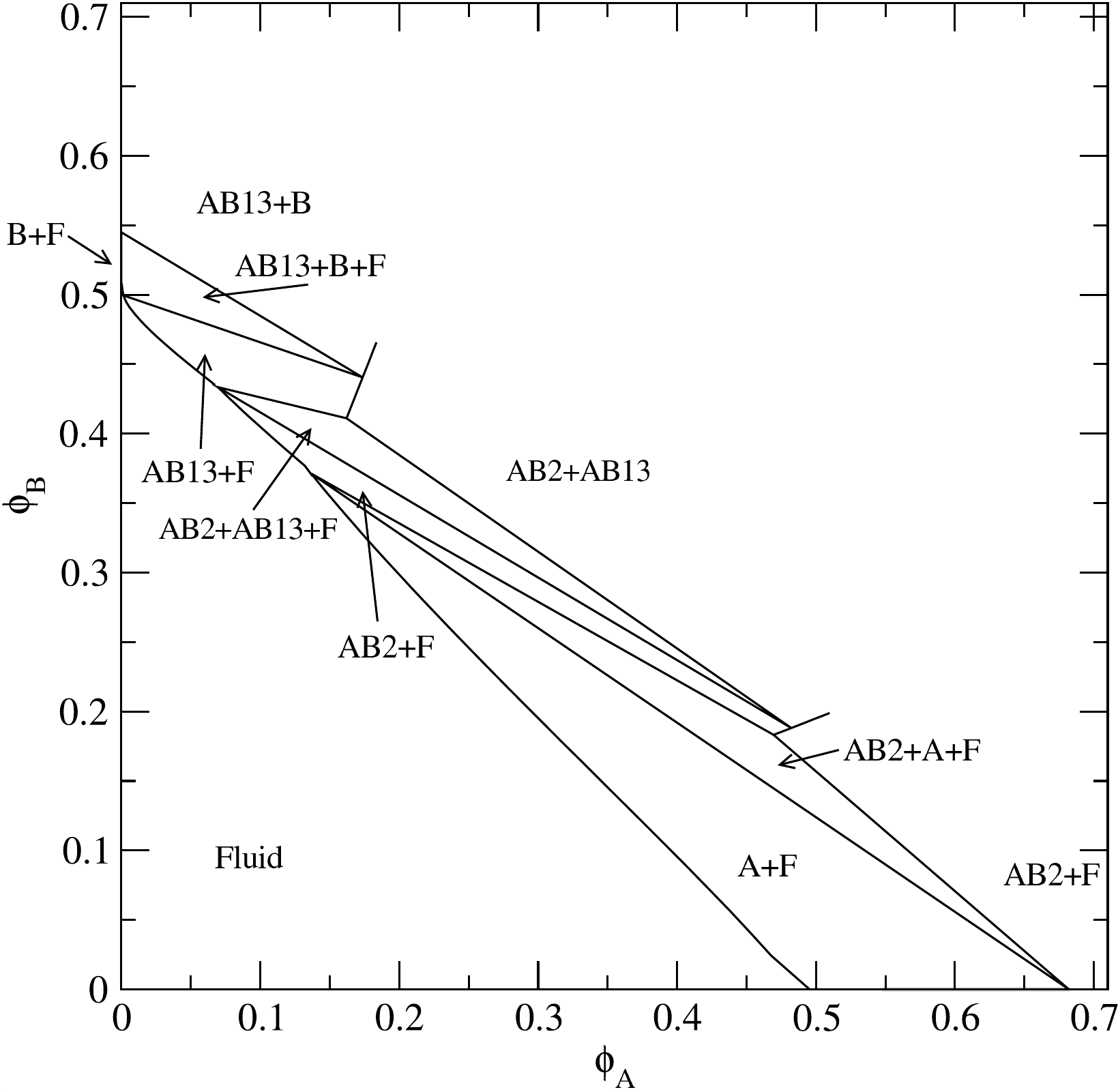}

\caption{\label{fig:VFPhases}Partial volume fraction phase diagram, built
from the P-X results shown in Figure \ref{fig:PPhases} using the
equation of state data to determine the densities of coexisting phases
along each P-X coexistence line.}
\end{figure}

Following the prescription outlined in section \ref{sub:Simulation-Methodology},
the lattice switch data were used to predict the pressure-composition
phase diagram for the binary hard sphere system at $\alpha=0.58$,
and the results are shown in Figure \ref{fig:PPhases}. The corresponding
partial volume-fraction phase diagram is shown in Figure \ref{fig:VFPhases}.
Both these figures agree well with the established results \cite{ELDRIDGE1993,COTTIN1995a},
although the partial volume fraction phase diagram does not extend
to as high densities as the one published in \cite{ELDRIDGE1993}
due to the limited range of pressures explored in the current work.

\section{Discussion}

We have presented a method by which the lattice switch approach can
be applied to binary mixtures, and applied it to binary hard sphere
system. The technical innovation introduced here involves including
a rescaling of particle sizes in the switch. Combining our free energy
differences enables us to build a phase diagram which is in excellent
agreement with previous theoretical results at the size ratio 0.58.
The only errors in the lattice-switch MC approach come from finite
size effects and sampling errors, which are easy to quantify. The
phase diagrams themselves rely on the Alder and Mansoori equations
of state for the FCC and liquid free energies respectively.

We also investigated the experimentally reported CsCl phase at size
ratio 0.73, and found it to be unstable compared to phase-segregated
FCC or the liquid. However the energy differences are small, and it
is possible that either charge effects or polydispersity may be responsible
for the experimental observation of this structure.

The dependence upon the Alder crystal is perhaps unfortunate, as it
is unclear how accurate the parameters of this equation of state are.
However, the Alder equation of state is dominated by the leading term
and rather insensitive the poorly-determined parameters. Our results
are in good agreement with the literature, including those \cite{ELDRIDGE1993}
which do not depend on the Alder formulation (due to using an Einstein
crystal as a reference state). In the same vein, building a lattice-switch
mapping to an analytically soluble reference crystal would also allow
the use of the Alder equation of state to be avoided. This Hamiltonian-switch
approach could map from an Einstein crystal or single-occupancy cell
model to the target potential and structure, and would allow absolute
free energies to be measured directly.

Other future work includes lower size ratio, although as mentioned
before, the mobile smaller spheres will present challenges, although
it may be possible to use tethers, in a manner similar to that employed
for the fluid-solid transition \cite{Wilding2002}. Also, these structures
have been observed for noble gas mixtures \cite{Layer2006}, and this
could be investigated by generalising this approach in the same manner
as the extension to soft-potentials of the original lattice switch
approach \cite{Jackson:2002aa}.

\begin{acknowledgments}
This work was supported by EPSRC under grants GR/S10377/01 and GR/T11753/01.
A.N.J. would like to thank Mike Cates for helpful comments and discussions. 
\end{acknowledgments}
\bibliographystyle{unsrt}
\bibliography{refs}

\begin{thebibliography}{10}

\bibitem{K1609:Keplar}
J~Keplar.
\newblock Strena sue de nuive sexangula.
\newblock 1611.

\bibitem{H1998:KepSol}
TC~Hales and SP~Ferguson.
\newblock The kepler conjecture, 1998.
\newblock Details of Keplar's conjecture and it's solution can be found at
  \hreflit{http://www.math.lsa.umich.edu/~hales/countdown/}.

\bibitem{ADB1997:lattice-switch}
AD~Bruce, NB~Wilding, and GJ~Ackland.
\newblock Free energy of crystalline solids: A lattice-switch monte carlo
  method.
\newblock {\em Physical Review Letters}, 79(16):3002--3005, 1997.
\newblock \hrefsoton{cond-mat/9706154}.

\bibitem{FRENKEL1984}
D~Frenkel and AJC Ladd.
\newblock New monte-carlo method to compute the free-energy of arbitrary solids
  - application to the fcc and hcp phases of hard-spheres.
\newblock {\em Journal Of Chemical Physics}, 81(7):3188--3193, 1984.

\bibitem{ELDRIDGE1993}
MD~Eldridge, PA~Madden, and D~Frenkel.
\newblock Entropy-driven formation of a superlattice in a hard-sphere binary
  mixture.
\newblock {\em Nature}, 365(6441):35--37, 1993.

\bibitem{ELDRIDGE1993b}
M.~Eldridge, P.~Madden, and D~Frenkel.
\newblock A computer-simulation investigation into the stability of the {A}{B}2
  superlattice in a binary hard-sphere system.
\newblock {\em Molecular Physics}, 80(4):987--995, 1993.

\bibitem{Eldridge:1993aa}
M.~Eldridge, P.~Madden, and D~Frenkel.
\newblock The stability of the {A}{B}$_{13}$ crystal in a binary hard sphere
  system.
\newblock {\em Molecular Physics}, 79(1):105--120, 1993.

\bibitem{ABD2000:LSMCmethod}
AD~Bruce, AN~Jackson, GJ~Ackland, and NB~Wilding.
\newblock Lattice-switch monte carlo method.
\newblock {\em Physical Review E}, 61(1):906--919, Jan 2000.
\newblock \hrefsoton{cond-mat/9910330}.

\bibitem{NBW2000:phaseswitch}
NB~Wilding and AD~Bruce.
\newblock Freezing by monte carlo phase-switch.
\newblock {\em Physical Review Letters}, 85(24):5138--5141, 2000.
\newblock \hrefsoton{cond-mat/0009062}.

\bibitem{Wilding2002}
N~Wilding.
\newblock A new simulation approach to the freezing transition.
\newblock {\em Computer Physics Communications}, 146(1):99--106, 2002.

\bibitem{Jackson:2002aa}
A~N Jackson, A~D Bruce, and G~J Ackland.
\newblock Lattice-switch monte carlo method: application to soft potentials.
\newblock {\em Physical review E, Statistical, nonlinear, and soft matter
  physics}, 65(3 Pt 2B):036710, Mar 2002.

\bibitem{errington:3130}
Jeffrey Errington.
\newblock Solid--liquid phase coexistence of the lennard-jones system through
  phase-switch monte carlo simulation.
\newblock {\em The Journal of Chemical Physics}, 120(7):3130--3141, 2004.

\bibitem{McNeil-Watson2006}
G~McNeil-Watson and N~Wilding.
\newblock Freezing line of the lennard-jones fluid: A phase switch monte carlo
  study.
\newblock {\em Journal Of Chemical Physics}, 124(6):064504, 2006.

\bibitem{Barrat:1986aa}
JL~Barrat, M~Baus, and JP~Hansen.
\newblock Density-functional theory of freezing of hard-sphere mixtures into
  substitutional solid solutions.
\newblock {\em Phys Rev Lett}, 56(10):1063--1065, Mar 1986.

\bibitem{0022-3719-20-10-011}
J~Barrat, M~Baus, and J~Hansen.
\newblock Freezing of binary hard-sphere mixtures into disordered crystals: a
  density functional approach.
\newblock {\em Journal of Physics C: Solid State Physics}, 20(10):1413--1430,
  1987.

\bibitem{Sanders1980}
J.V Sanders.
\newblock Close-packed structures of spheres of two different sizes. i.
  observations on natural opal.
\newblock {\em Philosophical Magazine A (Physics of Condensed Matter, Defects
  and Mechanical Properties)}, 42(6):705--720, Dec 1980.

\bibitem{Murray1980}
M.J Murray and J.V Sanders.
\newblock Close-packed structures of spheres of two different sizes. ii. the
  packing densities of likely arrangements.
\newblock {\em Philosophical Magazine A (Physics of Condensed Matter, Defects
  and Mechanical Properties)}, 42(6):721--740, Dec 1980.

\bibitem{BARTLETT1990}
P~Bartlett, RH~Ottewill, and PN~Pusey.
\newblock Freezing of binary-mixtures of colloidal hard-spheres.
\newblock {\em Journal Of Chemical Physics}, 93(2):1299--1312, 1990.

\bibitem{PUSEY1994}
P~Pusey.
\newblock Phase-behavior and structure of colloidal suspensions.
\newblock {\em Journal Of Physics-Condensed Matter}, 6:A29--A36, 1994.

\bibitem{ELDRIDGE1995}
MD~Eldridge, PA~Madden, PN~Pusey, and P~Bartlett.
\newblock Binary hard-sphere mixtures: a comparison between computer-simulation
  and experiment.
\newblock {\em Molecular Physics}, 84(2):395--420, 1995.

\bibitem{Hunt2000}
N~Hunt, R~Jardine, and P~Bartlett.
\newblock Superlattice formation in mixtures of hard-sphere colloids.
\newblock {\em Physical Review E}, 62(1):900--913, 2000.

\bibitem{Schofield2005}
A~Schofield, P~Pusey, and P~Radcliffe.
\newblock Stability of the binary colloidal crystals ab(2) and ab(13).
\newblock {\em Physical Review E}, 72(3):031407, 2005.

\bibitem{Layer2006}
M~Layer, A~Netsch, M~Heitz, J~Meier, and S~Hunklinger.
\newblock Mixing behavior and structural formation of quench-condensed binary
  mixtures of solid noble gases.
\newblock {\em Physical Review B}, 73(18):184116, 2006.

\bibitem{Redl:2003aa}
F~X Redl, K-S Cho, C~B Murray, and S~O'Brien.
\newblock Three-dimensional binary superlattices of magnetic nanocrystals and
  semiconductor quantum dots.
\newblock {\em Nature}, 423(6943):968--71, Jun 2003.

\bibitem{Royall2006}
C~Royall, M~Leunissen, A~Hynninen, M~Dijkstra, and A~van Blaaderen.
\newblock Re-entrant melting and freezing in a model system of charged
  colloids.
\newblock {\em Journal Of Chemical Physics}, 124(24):244706, 2006.

\bibitem{Abbas:2006aa}
Sayeed Abbas and Timothy~P Lodge.
\newblock Superlattice formation in a binary mixture of block copolymer
  micelles.
\newblock {\em Phys Rev Lett}, 97(9):097803, Sep 2006.

\bibitem{BARTLETT1990a}
P~Bartlett.
\newblock A model for the freezing of binary colloidal hard-spheres.
\newblock {\em Journal Of Physics-Condensed Matter}, 2(22):4979--4989, 1990.

\bibitem{COTTIN1995}
X~Cottin and PA~Monson.
\newblock A theory of solid-solutions and solid-fluid equilibria for mixtures.
\newblock {\em International Journal Of Thermophysics}, 16(3):733--741, 1995.

\bibitem{COTTIN1995a}
X~Cottin and PA~Monson.
\newblock Substitutionally ordered solid-solutions of hard-spheres.
\newblock {\em Journal Of Chemical Physics}, 102(8):3354--3360, 1995.

\bibitem{XU1992}
H~Xu.
\newblock A density functional-study of superlattice formation in binary
  hard-sphere mixtures.
\newblock {\em Journal Of Physics-Condensed Matter}, 4(50):L663--L668, 1992.

\bibitem{DENTON1990}
A~Denton.
\newblock Weighted-density-functional theory of nonuniform fluid mixtures -
  application to freezing of binary hard-sphere mixtures.
\newblock {\em Physical Review A}, 42(12):7312--7329, 1990.

\bibitem{SMITHLINE1987}
S~Smithline.
\newblock Density functional theory for the freezing of 1-1 hard-sphere
  mixtures.
\newblock {\em Journal Of Chemical Physics}, 86(11):6486--6494, 1987.

\bibitem{Schofield:2001bv}
A~Schofield.
\newblock Binary hard-sphere crystals with the cesium chloride structure.
\newblock {\em Physical Review E}, 64(5), 2001.

\bibitem{Leunissen2005}
M~Leunissen, C~Christova, A~Hynninen, C~Royall, A~Campbell, A~Imhof,
  M~Dijkstra, R~van Roij, and A~van Blaaderen.
\newblock Ionic colloidal crystals of oppositely charged particles.
\newblock {\em Nature}, 437(7056):235--240, 2005.

\bibitem{Bartlett2005}
P~Bartlett and A~Campbell.
\newblock Three-dimensional binary superlattices of oppositely charged
  colloids.
\newblock {\em Physical Review Letters}, 95(12):128302, 2005.

\bibitem{Hynninen2006a}
A~Hynninen, M~Leunissen, A~van Blaaderen, and M~Dijkstra.
\newblock Cuau structure in the restricted primitive model and oppositely
  charged colloids.
\newblock {\em Physical Review Letters}, 96(1):018303, 2006.

\bibitem{Trizac1997}
E~Trizac, M~Eldridge, and P~Madden.
\newblock Stability of the ab crystal for asymmetric binary hard sphere
  mixtures.
\newblock {\em Molecular Physics}, 90(4):675--678, 1997.

\bibitem{Velasco1999}
E~Velasco, G~Navascues, and L~Mederos.
\newblock Phase behavior of binary hard-sphere mixtures from perturbation
  theory.
\newblock {\em Physical Review E}, 60(3):3158--3164, 1999.

\bibitem{Bowles:1994aa}
Richard Bowles and Robin Speedy.
\newblock Cavities in the hard sphere crystal and fluid.
\newblock {\em Molecular Physics}, 83(1):113--125, 1994.

\bibitem{ALDER1964}
B~Alder.
\newblock Studies in molecular dynamics. iii. mixture of hard spheres.
\newblock {\em Journal Of Chemical Physics}, 40(9):2724, 1964.

\bibitem{BJA1968:MDfccvhcpHS}
BJ~Alder, WG~Hoover, and DA~Young.
\newblock Studies in molecular dynamics. v. high-density equation of state and
  entropy for hard-discs and spheres.
\newblock {\em Journal of Chemical Physics}, 49(8):3688--3696, Oct 1968.

\bibitem{YOUNG1979}
D~Young.
\newblock Studies in molecular-dynamics .17. phase-diagrams for step potentials
  in 2 and 3 dimensions.
\newblock {\em Journal Of Chemical Physics}, 70(1):473--481, 1979.

\bibitem{MANSOORI1971}
G~Mansoori.
\newblock Equilibrium thermodynamic properties of misture of hard spheres.
\newblock {\em Journal Of Chemical Physics}, 54(4):1523, 1971.

\bibitem{BAB1999:mcmcrev}
BA~Berg.
\newblock Introduction to multicanonical monte carlo simulations.
\newblock {\em Fields Institute Communications}, 1999.
\newblock \hrefsoton{cond-mat/9909236}.

\bibitem{Smith:1996aa}
GR~Smith and AD~Bruce.
\newblock Multicanonical monte carlo study of solid-solid phase coexistence in
  a model colloid.
\newblock {\em Phys Rev E Stat Phys Plasmas Fluids Relat Interdiscip Topics},
  53(6):6530--6543, Jun 1996.

\bibitem{Jackson2001}
A.N Jackson.
\newblock {\em Structural Phase Behaviour Via Monte Carlo Techniques}.
\newblock PhD thesis, 2001.

\bibitem{1953NM:MC}
N~Metropolis, A~Rosenbluth, M~Rosenbluth, AH~Teller, and E~Teller.
\newblock Equations of state calculations by fast computing machines.
\newblock {\em Journal of Chemical Physics}, 21:1087, 1953.

\bibitem{272995}
Makoto Matsumoto and Takuji Nishimura.
\newblock Mersenne twister: a 623-dimensionally equidistributed uniform
  pseudo-random number generator.
\newblock {\em ACM Trans. Model. Comput. Simul.}, 8(1):3--30, 1998.

\bibitem{Houle2003}
Paul Houle.
\newblock Rngpack 1.1a, http://www.honeylocust.com/rngpack/.

\end{thebibliography}

\end{document}